\documentclass[%
 reprint,
superscriptaddress,
 amsmath,amssymb,
 aps,
prb,
]{revtex4-2}

\usepackage{graphicx}% Include figure files
\usepackage{dcolumn}% Align table columns on decimal point
\usepackage{bm}% bold math
\usepackage{hyperref}% add hypertext capabilities
\hypersetup{colorlinks=true, citecolor=blue, urlcolor=blue, linkcolor=blue}

\usepackage{xcolor}
\usepackage{soul}
\usepackage[utf8]{inputenc}
\DeclareUnicodeCharacter{2212}{\textendash}
\newcommand{\ANF}{Cs$_2$(Ag,Na)FeCl$_6$}  
\newcommand{\AF}{Cs$_2$AgFeCl$_6$}  
\newcommand{\NF}{Cs$_2$NaFeCl$_6$}  
\newcommand{\FeIII}{Fe$^{3+}$} 
\newcommand{\Ueff}{U$_{eff}$}  
\begin{document}

\title{Electronic structure of the magnetic halide double perovskites \texorpdfstring{Cs$_2$(Ag,Na)FeCl$_6$} from first-principles}% Force line breaks with \\
\author{Johan Klarbring}
 \email{johan.klarbring@liu.se}
 \affiliation{% 
Theoretical Physics Division, \\
Department of Physics, Chemistry and Biology (IFM),
Link\"{o}ping University, SE\textendash581 83, Link\"{o}ping, Sweden
}%
\author{Utkarsh Singh}
 \affiliation{% 
Theoretical Physics Division, \\
Department of Physics, Chemistry and Biology (IFM),
Link\"{o}ping University, SE\textendash581 83, Link\"{o}ping, Sweden
}%
\author{Sergei I. Simak}
\affiliation{% 
Theoretical Physics Division, \\
Department of Physics, Chemistry and Biology (IFM),
Link\"{o}ping University, SE\textendash581 83, Link\"{o}ping, Sweden
}%
\affiliation{% 
Department of Physics and Astronomy,
\\
Uppsala University, SE\textendash75120 Uppsala, Sweden
}%
\author{Igor A. Abrikosov}
\affiliation{% 
Theoretical Physics Division, \\
Department of Physics, Chemistry and Biology (IFM),
Link\"{o}ping University, SE\textendash581 83, Link\"{o}ping, Sweden
}%

\date{\today}% It is always \today, today,

\begin{abstract}
A family of magnetic halide double perovskites (HDPs) have recently attracted attention due to their potential to broaden application areas of halide double perovskites into e.g. spintronics. Up to date the theoretical modelling of these systems have relied on primitive approximations to the density functional theory (DFT). In this paper, we study structural, electronic and magnetic properties of the \FeIII-containing HDPs \AF\ and \NF\ using a combination of more advanced DFT-based methods, including DFT+U, hybrid-DFT and treatments of various magnetic states. We examine the effect of varying the effective Hubbard parameter, U$_{eff}$, in DFT+U and the mixing-parameter, $\alpha$, in hybrid DFT on the electronic structure and structural properties. Our results reveal a set of localized Fe(d) states that are highly sensitive to these parameters. \AF\ and \NF\ are both antiferromagnets with Ne\'el temperatures well below room temperature and are thus in their paramagnetic (PM) state at the external conditions relevant to most applications. Therefore, we have examined the effect of disordered magnetism on the electronic structure of these systems and find that while \NF\ is largely unaffected, \AF\ shows significant renormalization of its electronic band structure. 

\end{abstract}

\maketitle

%\tableofcontents

\section{\label{sec:intro} Introduction}

Research into metal halide perovskite (MHP) semiconductors have steadily increased over the last decade to where it is currently one of the most active fields in materials science \cite{Jena2019}. While lead-containing MHPs have been at the forefront of this research, a number of lead-free derivatives have emerged as interesting alternatives \cite{Ning2019_2}. These sub-classes often aim to address deficiencies, not just related to the toxicity of lead, but also in terms of poor stability of several MHPs. The so-called lead-free halide double perovskites (HDPs), where the Pb$^{2+}$ ions are replaced with a pair of mono- and trivalent cations, is one such class of materials which has emerged over the last years \cite{Fermi2019}. 

Very recently, a set of \FeIII\-based HDPs, including \NF\ and \AF, have garnered attention \cite{yin2020,Li2021,Ji2021}. While \NF\ was synthesized at least as early as 1976 \cite{Pebler1976}, renewed attention has been gained just in the last few years. In addition to \NF\ and \AF, related alloys have also been synthesized \cite{Ji2021}. These systems contain magnetic \FeIII\ ions with a 3d$^{5}$ electronic configuration and initial experiments indicate that they are in their high-spin (HS) configuration, yielding a large magnetic moment \cite{Xue2022}. As such, an associated set of physical phenomena related to their magnetism, become of relevance. \NF\ and \AF\ both crystallize in the ideal cubic double perovskite structure with space group Fm$\overline{3}$m (No.\ 225) at ambient conditions. They are further found to be antiferromagnets (AFM) with low N\'{e}el temperatures of ~3 K and ~18 K, respectively \cite{Xue2022}. As such, they are both deep in their paramagnetic (PM) states at ambient conditions. 

The presence of the partially filled 3d shell differentiates \NF\ and \AF\ from the majority of MHPs studied in the literature. These Fe(3d) states are expected to be highly localized and an accurate treatment typically requires consideration beyond standard semi-local DFT. In this work, we explore two such approaches. The first is the DFT+U method \cite{Himmetoglu2014}, where an (effective) onsite Hubbard term, \Ueff, is added to the effective Kohn-Sham potential of the localized states. This method is advantageous in that it adds negligible computational effort compared to standard semi-local DFT. However, a choice needs to be made on the value of the effective U$_{eff}$ parameter. While this can in principle be estimated from first principles \cite{Cococcioni2005}, often a value is taken to match some experimentally determined property. The second approach is hybrid DFT, where a fraction of exact exchange is mixed in with the semi-local DFT exchange. This approach is often chosen for semiconductors and insulators when accurate electronic structures are desired. Although it acts in a similar way to DFT+U on highly localized states\cite{Ivady2014}, it is not as popular for studies of such systems. Computationally, hybrid DFT is significantly more expensive than DFT+U and suffers from a similar problem as DFT+U when it comes to choosing the amount, $\alpha$, of exact exchange to mix in. 

A second challenge in modelling these systems, also rarely encountered for MHPs, is that they are in the paramagnetic (PM) state at the relevant external conditions. The PM state is sometimes modeled as non-magnetic, without the presence of local magnetic moments, which could lead to erroneous conclusions \cite{Alling2010_2}.

Previous DFT studies on \NF\ and \AF\ \cite{Xian2020,yin2020,Radja2022} have either been performed with a NM or FM state and have not directly addressed the issue of the modelling of the strongly localized Fe(d) states. In the current paper, we therefore aim to explore how DFT+U and hybrid-DFT behave in modelling these systems and study the influence of varying values of U and the mixing parameter, $\alpha$, respectively. Moreover, we use a description of the PM state in the framework of the disordered local moment (DLM) picture \cite{Gyorffy1985} using a DLM-supercell approach \cite{Alling2010}, to explore how the disordered PM state influences structural and electronic structure properties.

\section{\label{sec:comp_details}Computational details}

All computations were performed using plane-wave Kohn-Sham DFT and the PAW method \cite{blochl1994} as implemented in VASP \cite{kresse1996,kresse1996_2,kresse1999}. We have used the PBEsol \cite{perdew2008} form of the exchange-correlation functional augmented with an effective on-site U$_{eff}$ correction in the form according to Dudarev et al.\cite{Dudarev1998}, as well as range-separated hybrid DFT \cite{Krukau2006} with varying values of the mixing-parameter $\alpha$. For the majority of the calculations we have used a 450 eV cutoff energy, and an 8$\times$8$\times$8 sampling of the first Brillouin zone (BZ) of the 10 atom primitive unit cell and a 10$^{-6}$ eV convergence criterion for the self-consistent field (SCF) cycles and the "\texttt{PREC = Accurate}" mode in VASP. Full structural relaxations (lattice vectors and internal coordinates) were performed until all the atomic forces were $<$ 5 meV/Å. For relaxations with hybrid-DFT we have used a less dense 4$\times$4$\times$4 k-point grid. We confirmed that yielded forces $<$ 6 meV/Å different compared with the 8$\times$8$\times$8 grid used for calculations of densities of states (DOS). The employed PAW potentials treated the Cs(5s5p6s), Ag(4d5s), Na(2p3s), Fe(3p3d4s) and Cl(3s3p) states as valence. 
Details on chemical bonding was obtained from a Crystal Orbital Hamiltonian Population (COHP) \cite{Dronskowski1993} analysis, using the LOBSTER\cite{LOBSTER1,LOBSTER2,LOBTER3} package.

The PM state was modelled with special quasirandom structure (SQS) \cite{Zunger1990} distributions of the spin-up and spin-down Fe magnetic moments in a disordered-local moment (DLM) fashion \cite{Alling2010}. For the calculations of energies of different magnetic states we used a 320 atom supercell, constructed as a 2$\times$2$\times$2 repetition of the 40 atom conventional cubic unit cell, and $\Gamma$-point sampling of the BZ. Effective band structures (EBS) in the BZ of the primitive unit cell were obtained using band unfolding \cite{Popescu2012} as implemented in the BandUP code \cite{Medeiros2014}. For these calculations we used a 4$\times$4$\times$4 supercell of the primitive fcc unit cell (640 atoms). In order to make these calculations computationally feasible, the accuracy settings were slightly reduced as follows: cutoff energy 300 eV, energy convergence criterion 10$^{-5}$ eV and "\texttt{PREC = Normal}". We have recalculated the band structures of the FM unit cell with these settings to confirm that band structures agree almost within the width of the line in the displayed band structures.

\section{\label{sec:results}Results}
\begin{figure*}[t]
\centering
\includegraphics[width=\linewidth]{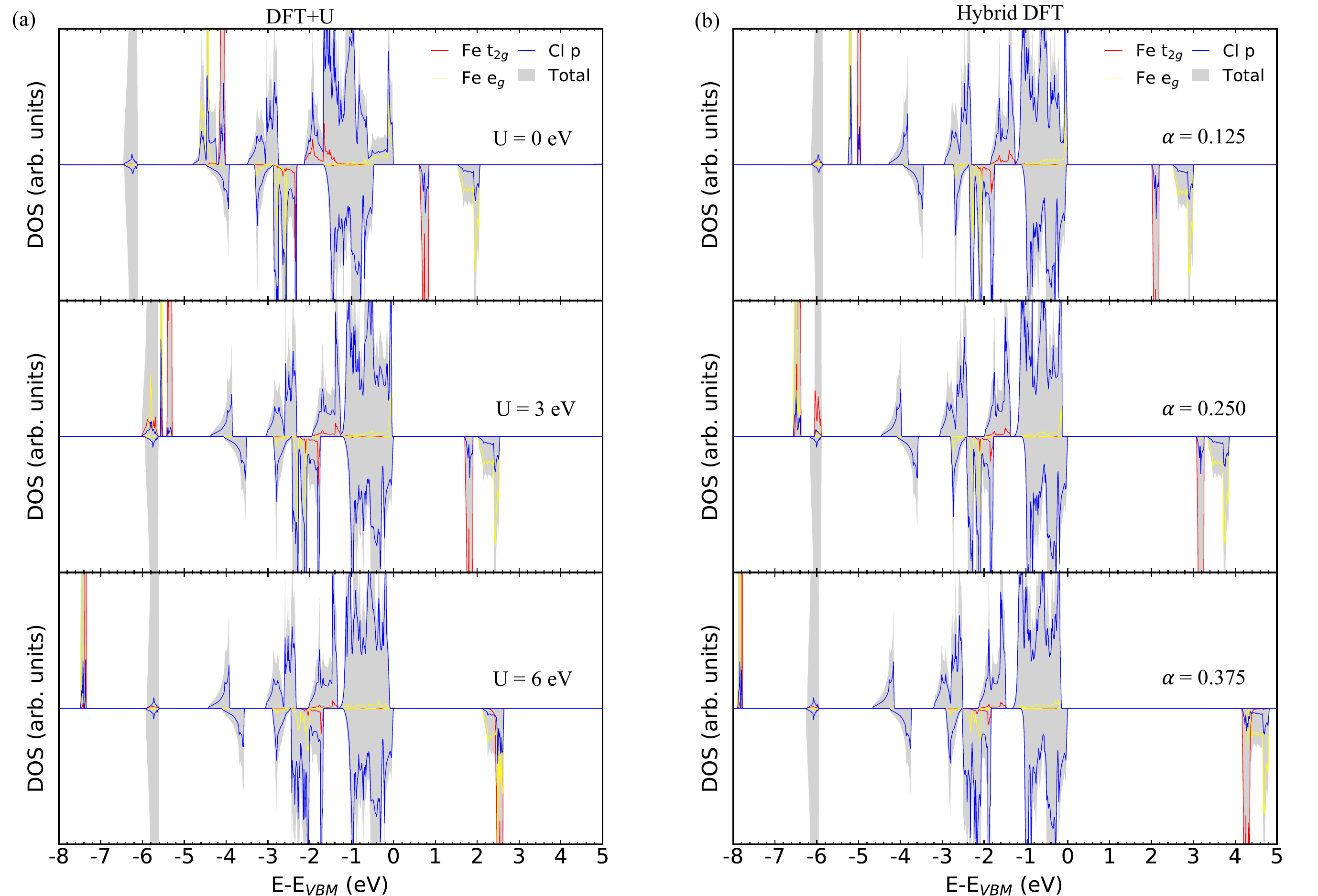}
\caption{Electronic DOS for \NF\ in the FM state, for (a) varying values of U$_{eff}$ on the Fe(d) states in the PBEsol+U methodology and (b) varying values of $\alpha$ in the hybrid DFT (default HSE06 range separation parameter). The DOS are aligned to the VBM.}
\label{fig:DOS_Na}
\end{figure*}
\begin{figure}[t!]
\centering
\includegraphics[width=\linewidth]{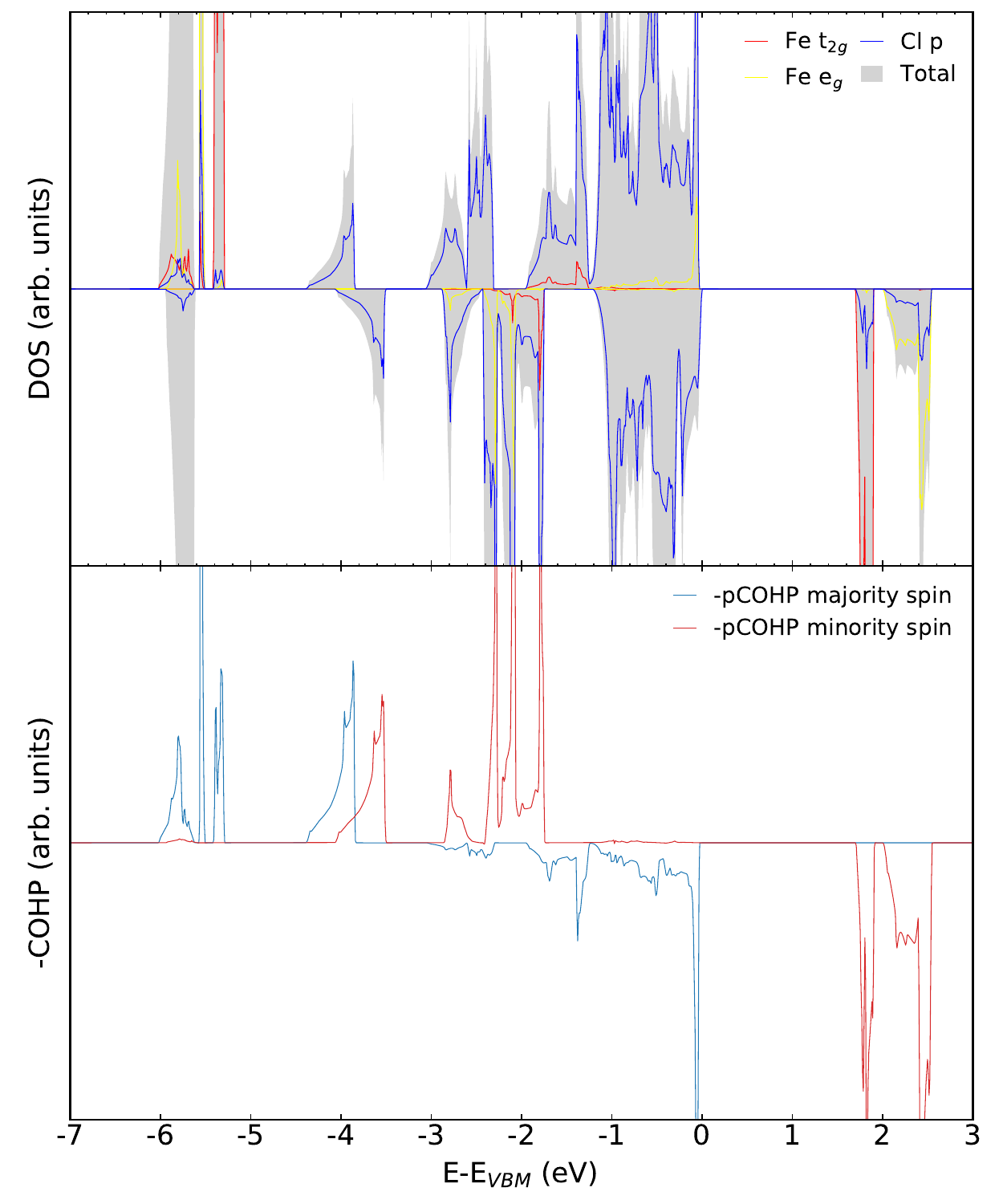}
\caption{(a) PBEsol+U(3eV) density of states and (b) corresponding Fe-Cl partial COHP for the majority (blue line) and minority (red line) spin-channels.The bonding states around -4 eV are of Fe(s)-Cl(p) character. }
\label{fig:COHP}
\end{figure}

\begin{figure}[t!]
\centering
\includegraphics[width=\linewidth]{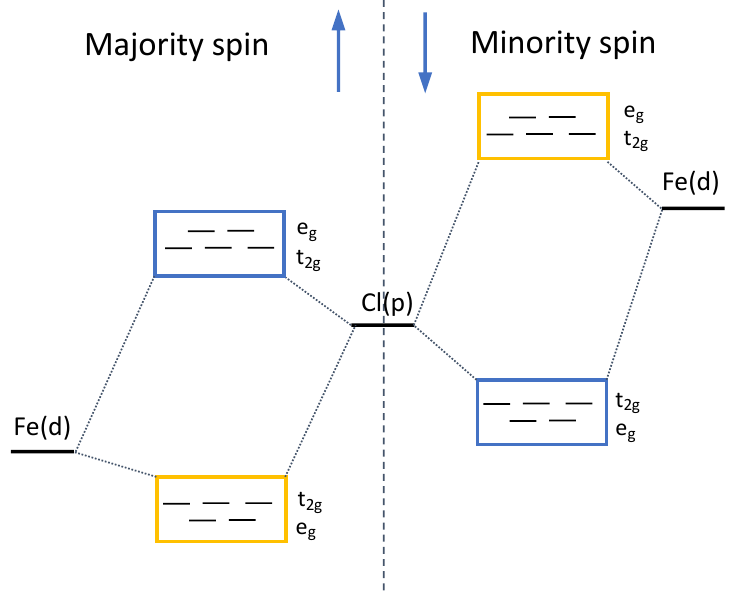}
\caption{Schematic cartoon illustration of Fe(d)-Cl(p) hybridization in \NF. Yellow and blue boxes indicate states of predominately Fe(d) and Cl(p) character, respectively.}
\label{fig:bonding_schematic}
\end{figure}

\begin{figure*}[t]
\centering
\includegraphics[width=\linewidth]{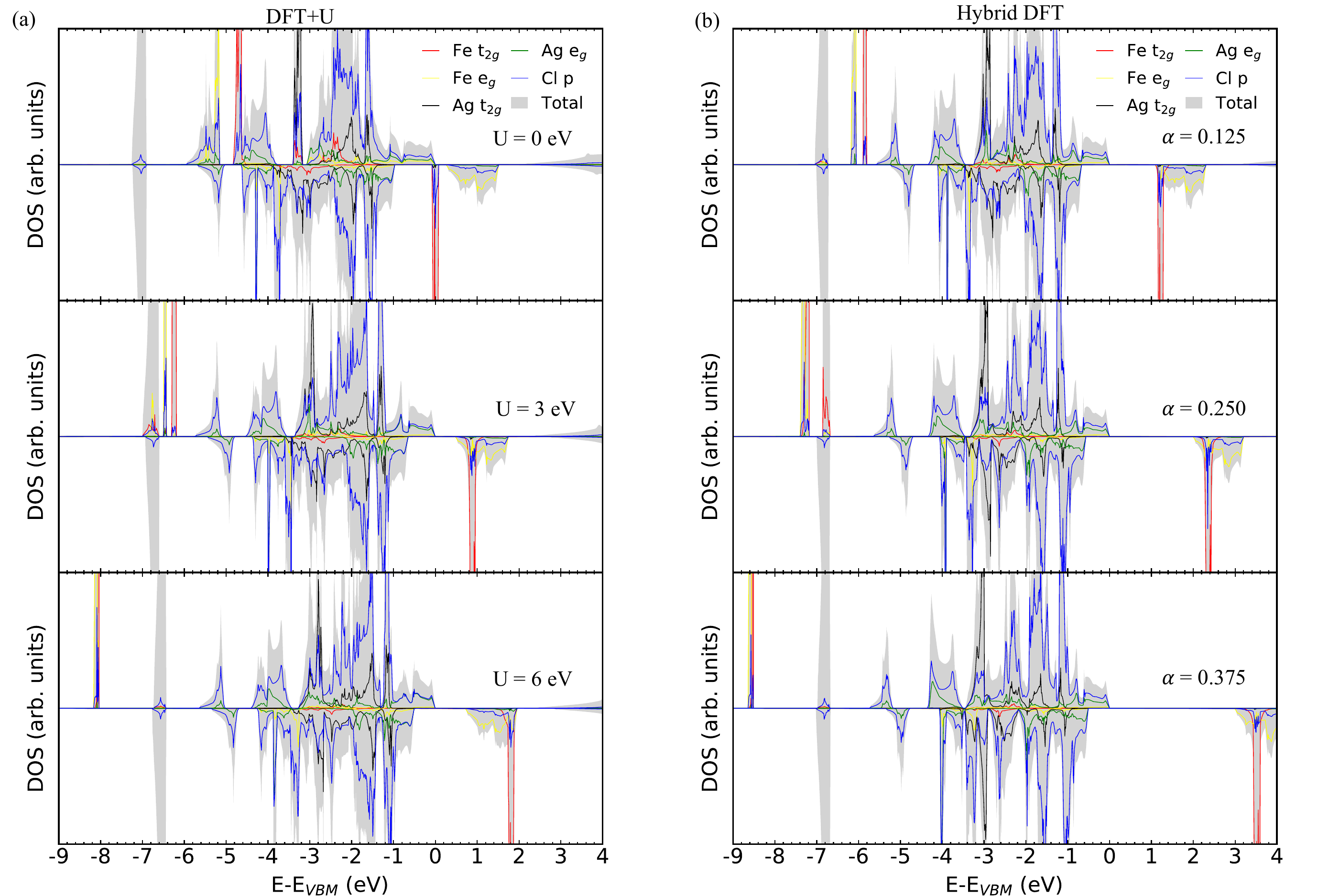}
\caption{Electronic DOS for \AF\ in the FM state, for (a) varying values of U on the Fe(d) states in the PBEsol+U methodology and (b) varying values of $\alpha$ in the hybrid DFT (default HSE06 range separation parameter). The DOS are aligned to the VBM.}
\label{fig:DOS_Ag}
\end{figure*}

\begin{figure}[t!]
\centering
\includegraphics[width=\linewidth]{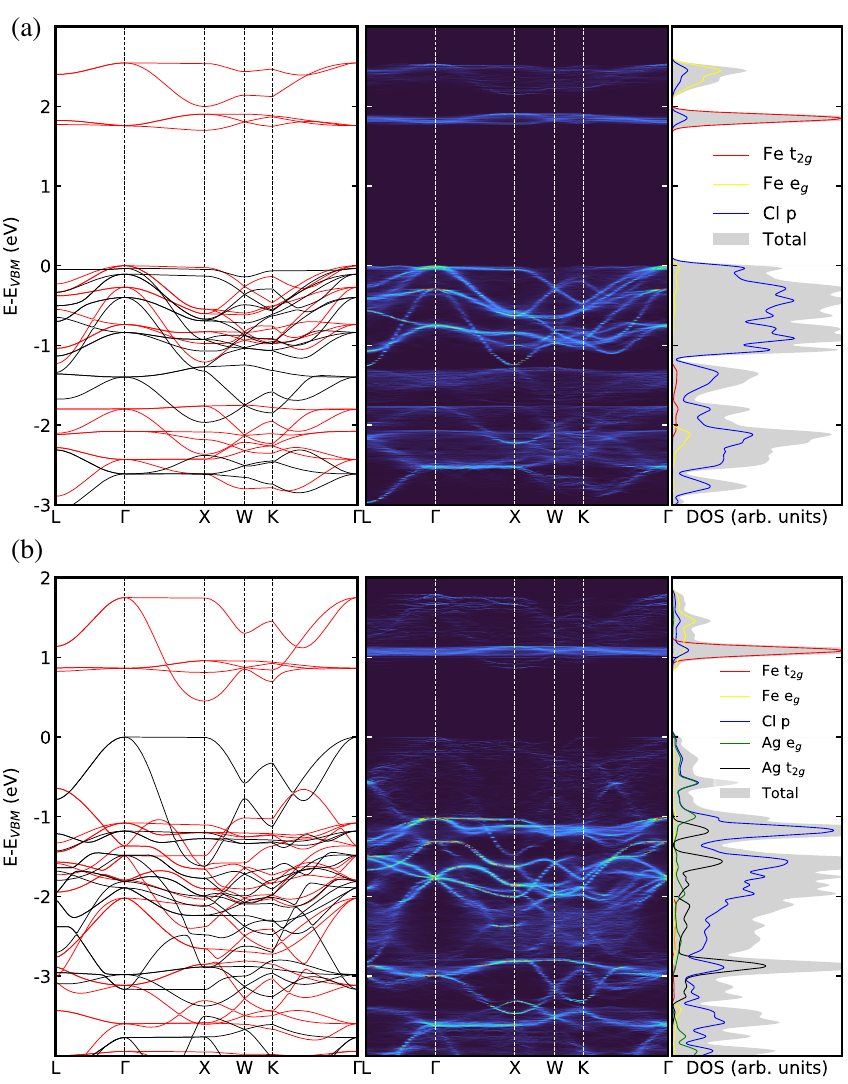}
\caption{Effective band structure (EBS) of (a) \NF\ and (b) \AF\ in DLM state. The left panels shows the bands structure in the FM state for comparison, with the majority and minority spin bands in black and red, respectively. The right panels is the DOS in the DLM states. High and low spectral weights are represented by bright and dark colors, respectively. }
\label{fig:EBS}
\end{figure}

\subsection{\label{sec:basics}Basic electronic and structural properties}
We start by examining the structural and magnetic properties, as well as the electronic structure of \NF\ and \AF\ across a range of U$_{eff}$ and $\alpha$ values. Although, as mentioned in the introduction, \NF\ and \AF\ are both AFM at low temperature and in their disordered PM state at relevant external conditions, we will in this section primarily treat them in their ferromagnetic (FM) state. This allows us to retain the symmetry of the double perovskite structure and to most easily highlight basic features of their electronic structure. The effect of different magnetic states on the electronic structure will then be considered in more detail in Section \ref{sec:magnetism}.

The double perovskite structure contains 2 degrees of freedom, the lattice parameter, $a$, (or equivalently the unit cell volume) and the position of the Cl atom along the (Ag/Na)-Cl-Fe bond. Table \ref{table:struct} lists these parameters for both \NF\ and \AF\ for different U values in PBEsol+U and different $\alpha$ values in the HSE methodology. To assess the importance of the magnetic state on the structural parameters, we show results for the FM, PM and NM states in the case of U$_{eff}$ = 3 eV.

Generally, PBEsol+U gives the better structural parameters in comparison to experiments, with HSE somewhat overestimating lattice constants. The structural parameters are rather weakly influenced by varying the values of U$_{eff}$ and $\alpha$. It is, however, interesting to note that increasing U$_{eff}$ increases the Fe-Cl bond length and lattice constant, while increasing $\alpha$ instead decreases the same bond length and lattice constant. We can also see from Table \ref{table:struct} that, while the NM calculations clearly gives erroneous results for the lattice constant and bond lengths, the precise nature of the magnetic state, i.e. FM or PM, is not influential on the structural parameters, as long as local magnetic moments are included in the calculations. 

\begin{table*}
\caption{\label{table:struct}Lattice constant, $a$ and select bond-lengths (in units of Å), total magnetic moments in the unit cell, $\mu$(uc), and local magnetic moments on the \FeIII-ion, $\mu$(Fe). For the PM states, the average values of the bond lengths and $\mu$(Fe) are given. Experimental results are from Refs.\ \cite{Li2021,yin2020} }
\begin{ruledtabular}
\begin{tabular}{ccccccc}
&  & \NF\ &   \\
\hline
Method & Magnetic state &a & d(Fe-Cl) & d(Na-Cl) & $\mu$(uc) & $\mu$(Fe) \\
\hline
Experiment (80 K) \cite{Li2021} & PM  & 10.2591 & 2.366 & 2.764  \\

\hline
  PBEsol+U \\   
  \hline
    U=0 eV  & FM & 10.246 & 2.377 & 2.746 & 5.00 & 3.88 \\
    \hline
    U=3 eV  & FM & 10.258 & 2.381 & 2.747 & 5.00 & 4.06 \\
            & NM & 10.073 & 2.276 & 2.761 & 0 & 0 \\
            & PM & 10.258 & 2.381 & 2.748 & 5.00 & 4.06 \\
    \hline
    U=6 eV  & FM & 10.266 & 2.385 & 2.748 & 5.00 & 4.22 \\
    \hline
    U=9 eV  & FM & 10.271 & 2.388 & 2.748 & 5.00 & 4.38 \\
\hline  
    HSE \\   
 $\alpha$ = 0.125 & FM & 10.461 & 2.407 & 2.823 & 5.00 & 4.04 \\
 $\alpha$ = 0.250 & FM & 10.432 & 2.399 & 2.817 & 5.00 & 4.16 \\
 $\alpha$ = 0.375 & FM & 10.397 & 2.392 & 2.807 & 5.00 & 4.26 \\

\hline  
& \\
 &  & \AF\ &   \\
\hline
Method & Magnetic state & a & d(Fe-Cl) & d(Na-Cl) & $\mu$(uc) & $\mu$(Fe) \\
\hline
Experiment (300 K) \cite{yin2020}& PM & 10.2023 & 2.382 & 2.719  \\
\hline
  PBEsol+U \\   
    U=0 eV  & FM & 10.079 & 2.387 & 2.652 & 4.53 & 3.67\\
    \hline
    U=3 eV  & FM & 10.126 & 2.396 & 2.667 & 5.00 & 4.01\\
            & NM & 9.93 & 2.295 & 2.668 & 0 & 0 \\
            & PM & 10.116 & 2.381 & 2.747 & 5.00 & 4.06 \\
    \hline
    U=6 eV  & FM & 10.139 & 2.400 & 2.669 & 5.00 & 4.16\\
    \hline
    U=9 eV  & FM & 10.150 & 2.402 & 2.673 & 5.00 & 4.32\\

\hline  
    HSE \\   
 $\alpha$ = 0.125 & FM & 10.377 & 2.416 & 2.773  & 5.00 & 4.01\\
 $\alpha$ = 0.250 & FM & 10.364 & 2.405 & 2.778  & 5.00 & 4.13\\
 $\alpha$ = 0.375 & FM & 10.348 & 2.395 & 2.779  & 5.00 & 4.24\\

\end{tabular}

\end{ruledtabular}
\end{table*}

Table \ref{table:struct} also shows the total magnetic moment in the unit cell and the local magnetic moments on the Fe atoms for different values of U$_{eff}$ and $\alpha$ \footnote{Using the default radius around the Fe atoms (\texttt{RWIGS}), as supplied by the VASP PAW potentials}. We see that in all cases, expect for U=0 eV in \AF\ there is a total magnetic moment of 5 $\mu_B$ in the unit cell, as expected from the 5 unpaired electrons of HS \FeIII. The local magnetic moment on the Fe ions increases for increasing U$_{eff}$ (or $\alpha$), due to the increased charge localization.

Turning to the electronic structure, Fig.\ \ref{fig:DOS_Na} shows the densities of states (DOS) in the FM spin configuration for \NF, with U$_{eff}$values on the Fe 3d states varying from 0 to 6 eV.

The most eye-catching feature of the DOS is the highly localized Fe-d states. These separate, as dictated by the octahedral coordination environment of the Cl-p ligands, into states with e$_{g}$ and t$_{2g}$ symmetry. In a simple picture, the e$_{g}$ and t$_{2g}$ orbitals have both Coulomb and covalent interactions with the ligands. The e$_{g}$ orbitals point more directly towards the ligands and the Coulomb repulsion is thus stronger than for the t$_{2g}$ orbitals, but the covalent interaction also tends to be larger. 

In general, increasing the value of U$_{eff}$ has the effect of driving the occupied and unoccupied states to which the U$_{eff}$ is applied, Fe 3d states in our case, down and up in energy, respectively. This can clearly be seen to be the case for \NF\ in Fig. \ref{fig:DOS_Na}. Importantly, however, the sharper t$_{2g}$ and the broader e$_{g}$ states move with U$_{eff}$ at different rates. This is particularly true for the unoccupied e$_{g}$/t$_{2g}$ states, where the t$_{2g}$ states are initially lower in energy, but where a crossing with the e$_{g}$ states happens at U$_{eff}$ $\approx 6$ eV. This is because the e$_{g}$ states have, as mentioned above, a stronger covalent interaction with the ligands and as such have a counteracting effect to the increase in U$_{eff}$, while the more localized t$_{2g}$ states move with U$_{eff}$ at will. 

There are two sets of Fe(d)-Cl(p) hybridized states in each spin-channel, one of predominantly Fe(d) character and one with predominantly Cl(p) character. In the majority spin-channel both sets are occupied, while the predominantly Fe(d) states are unoccupied in the minority spin-channel. We see that the predominantly Fe(d) peaks corresponding to occupied e$_{g}$ and t$_{2g}$ states and unoccupied t$_{2g}$ states are very sharp,  while the unoccupied e$_{g}$ band is more disperse. 

To get further insight into the electronic structure of \NF, Fig.\ \ref{fig:COHP} shows the Fe-d partial DOS for U$_{eff}$ = 3 eV, along with the Fe-Cl partial crystal orbital Hamiltonian population (COHP). Analysing the Fe(3d)-Cl(3p) hybridised states, we see that the predominantly Fe(d) peaks are of bonding and anti-bonding character in the majority and minority spin-channels, respectively, and that the order of the t$_{2g}$ and e$_{g}$ states is opposite in the two spin-channels. The predominantly Cl(p) states are instead anti-bonding in the majority and bonding in the minority spin-channel. As expected, the bonding-antibonding energy splitting is larger for the e$_{g}$ orbitals, resulting in the opposite order of t$_{2g}$/e$_{g}$ states for the bonding and antibonding Fe(3d)-Cl(3p) combinations, for lower values of U. 

Increasing the value of U$_{eff}$ has the additional effect of localizing the Fe-d states and thus reducing the covalent interaction with the ligands. The impact of this is seen most clearly for the Fe(3d)-Cl(3p) anti-bonding states in the majority spin-channel, where, upon increasing U$_{eff}$, the Fe(e$_{g}$/t$_{2g}$) weight rapidly decreases and gets redistributed into the sharp bonding states. In fact, the effect of adding U$_{eff}$ is rather drastic on the top of the valence band; for U$_{eff}$ = 0 eV, antibonding Fe(e$_{g}$)-Cl(p) states are separated out from the rest of the VB in the majority spin-channel. This is a clear indication that "regular" semilocal DFT functionals, such as PBEsol, do not accurately describe the electronic structure of \NF.

As can be seen in Fig. \ref{fig:DOS_Na} b), the effect of increasing the mixing parameter $\alpha$ has similar impact on the Fe(d) states as increasing U$_{eff}$. This is to be expected as DFT+U and hybrid DFT can be shown to behave very similarly for localized states \cite{Ivady2014}. An important difference, however, is that the unoccupied t$_{2g}$ states move up in energy on increasing U$_{eff}$ at a much larger rate than the e$_{g}$ states, this rate difference is much smaller on increasing $\alpha$. Indeed, the unoccupied t$_{2g}$ states stay below the e$_{g}$ states at least up to $\alpha$=0.5. This difference in behavior could be related to different spatial localization in hybrid DFT and DFT+U calculations resulting in somewhat different splitting between Coulomb and covalent interactions \cite{UHybrid}.

Another interesting observation is that the sharp occupied bonding Fe(e$_{g}$)-Cl(p) states are lower in energy than the t$_{2g}$ states. This is contrary to the predictions from simple crystal field theory (CFT) applied to an octahedral coordination environment, where the opposite ordering of t$_{2g}$ and e$_{g}$ states is expected. Indeed, if the atomic d orbitals are split simply by their Coloumb interaction with the ligands, represented as negative point charges, the e$_{g}$ orbitals have to lie above the t$_{2g}$ orbitals. This is, as mentioned above, because e$_{g}$ orbitals point more directly towards the ligands and are thus repelled more strongly. However, moving beyond CFT and considering also covalent interactions between the e$_{g}$/t$_{2g}$ orbitals and the ligands, things are a bit less straightforward.

For the typical case where the transition metal d orbitals lie above the ligand p levels, the anti-bonding combination of ligand p and TM d are the ones which are predominately of TM d-character and which are often refereed to as just the TM d-states. These states also show the "normal" order of e$_{g}$/t$_{2g}$ orbitals \cite{Ushakov2011}. However, in the opposite case where the TM d lie below the ligand p levels, the predominantly d levels are the bonding orbitals which can show the opposite t$_{2g}$/e$_{g}$ ordering. In the present case, we propose that the Fe(3d) and Cl(3p) hybridisation in \NF\ can be schematically illustrated as in Fig.\  \ref{fig:bonding_schematic}.

Fig. \ref{fig:DOS_Ag} shows the DOS for different U$_{eff}$ and $\alpha$ for \AF, which is, in many ways, similar to \NF. Indeed, the Fe(e$_{g}$/t$_{2g}$) states behave in much the same way, the primary difference being that the unoccupied e$_{g}$ band is broader in \AF\ than in \NF. This is due to a stronger Fe-Fe interaction as will be discussed further below. The main difference between the two systems, however, is the presence of Ag(d) states in the main valence band. In particular bands of predominantly Ag(e$_{g}$)-Cl(p) character are at the top of the valence band in the majority spin-channel. It is thus clear that beside the \FeIII -ions, also the Ag$^{+}$ ions are spin-polarized, despite the fact that the they have a full d$^{10}$ electron configuration. As we will see below, these Ag(e$_{g}$) states are important in mediating the magnetic interaction between \FeIII, and is responsible for the different magnetic energetics in \AF\ compared to \NF. 

It is worth noting that for U$_{eff}$= 0 eV, the states at the top of the valence band in the majority spin-channel overlap in energy with the sharp Fe(t$_{2g}$) peak in the minority spin-channel, resulting essentially in a metallic system.  

\subsection{\label{sec:magnetism} Magnetic states}

\begin{table}
\caption{\label{table:mag}Energies of different magnetic states, referenced to the FM configuration, in \NF\ and \AF.  }
\begin{ruledtabular}
\begin{tabular}{ccc}
&\multicolumn{2}{c}{$\Delta E$ (meV/f.u.)}     \\
\hline
Magnetic state &   \NF\ & \AF\ \\
\hline
FM & 0 & 0  \\
AFM-I & -8 & -56 \\
PM  & -6  & -49 \\
\end{tabular}

\end{ruledtabular}
\end{table}
While the initial investigation in the previous section was, for clarity and simplicity, done primarily with a FM spin-configuration of the Fe atoms, experiments indicate that both \NF\ and \AF\ have AFM ground-states with low Ne\'el temperatures \cite{Xue2022}. Table.\ \ref{table:mag} lists the total energies of the NM, FM, AFM-I \footnote{The AFM-I consists of spin-up and spin-down layers of Fe magnetic moments alternating in the z direction.} and PM spin configurations. The calculations used PBEsol+U(3eV), this \Ueff\  was chosen since it gives the same ordering of the unoccupied e$_{g}$/t$_{2g}$ states as hybrid DFT for reasonable values of $\alpha$. The relaxed FM lattice constant was used and internal degrees of freedom were relaxed for each magnetic state. 

We see that in both cases the AFM-I configuration is lowest in energy, in agreement with the experimental results that these systems are AFM at low temperatures. For \NF, however, the energies across all three spin configurations are close to degenerate on the $\sim$1 meV/atom accuracy in our calculations, indicating a weak interaction between Fe magnetic moments in this system. This is in agreement with the very low experimental Ne\'el temperature of $\sim$3 K.

For \AF, the energy differences across the spin configurations are larger, but the AFM-PM energy difference is still small, again in agreement with a low Ne\'el temperature of $\sim$18 K. The energy difference between the AFM/PM and the FM state is however much larger in \AF\ compared to \NF, indicating some significant interaction between Fe magnetic moments. This is likely due to the Ag d shell which can mediate the magnetic interactions between the Fe spins to a larger degree than the mostly inert Na ion. A detailed study into the magnetic interactions in magnetic halide double perovskites is beyond the scope of the present work and will be reported elsewhere.  

Since the Ne\'el temperatures of \ANF\ are low, the relevant magnetic setting for any prospective applications at or around ambient temperature is the PM phase. To model the electronic band structure in the PM state we employ a DLM spin-configuration, and then use the band unfolding formalism \cite{Popescu2012} to unfold the resulting band structure, yielding an effective band structure (EBS) in the primitive BZ of the double perovskite structure. The results are shown in Fig. \ref{fig:EBS} for \NF\ and \AF, where we compare these EBSs to their corresponding band structures in the FM state. 

We see that for \NF\ the band structure is rather weakly affected by the different local spin environments that the Fe atoms experience and the EBS resembles the FM band structure quite closely, with minor shifts and broadening in certain bands with some Fe(e$_{g}$/t$_{2g}$) character. 

The effects of the spin configuration on the band structure is significantly larger for \AF. Firstly, we see that the PBEsol+U(3eV) bandgap is significantly increased in the PM as compared to the FM spin configuration. This follows from the decreased bandwidth in the PM state of the unoccupied Fe e$_{g}$ band at the bottom of the conduction band and occupied Ag e$_{g}$ band at the top of the valence band in the $\Gamma-X$ direction. This is a result of the weaker Fe-Fe interaction in this direction upon introduction of magnetic disorder and is present, although to a lesser extent, also for the unoccupied Fe e$_{g}$ band in \NF. 

Indeed, for \AF\ all the states within $\sim$1 eV of the valence band maximum, of Ag(e$_{g}$)+Cl(p) character, are rather heavily broadened and shifted in the EBS as a result of disordered spin configuration on the Fe ions. This indicates that the bands of Ag(e$_{g}$) character are significantly influenced by the spin configuration of the neighboring Fe ions. This in turn suggests a stronger interaction between Fe spins, mediated by Ag(e$_{g}$) orbitals, which explains the larger energy difference between the FM and AFM spin configurations in \AF\ compared to \NF.

It is important to note at this point that broadening effects in the EBS may converge quite slowly with the size of the employed supercell \cite{Wang2021}. Unfortunately, the computational complexity of employing larger supercells than the 640 atom ones used in this work is prohibitively large. Nevertheless, the presented result should be qualitatively accurate in terms of which states are effected by the disordered magnetism.   

Analysing the band structures of \AF\ and \NF\ for U$_{eff}$ = 3 eV in their FM states, we find that for spin-allowed transitions \AF\ is an indirect bandgap semiconductor in the minority spin-channel, with the VBM and CBM at the $L$ and $X$-points, respectively. \NF\ is also an indirect semiconductor with VBM and CBM at the $\Gamma$ and $X$-points, respectively, due to the almost dispersionless $\Gamma-X$ branch at the top of the valence band, the direct bandgap is very close to the fundamental gap.

In the DLM state, where the two spin channels are equivalent, both \AF\ and \NF\ are practically direct band gap systems due to their flat $\Gamma$-$X$ branches in the VB and the CBM at the $X$-point.

\section{\label{sec:discussion}Discussion}
We now provide some thoughts regarding the choice of $\alpha$ and $U_{eff}$. As we have seen, basic structural properties are not strongly affected by these parameters, whereas the electronic structure is. It might therefore be tempting to attempt to tune U$_{eff}$ or $\alpha$ to match the room temperature experimental bandgap values, 2.07-2.2 eV \cite{Xian2020,Ji_manus} and 1.55-1.6 eV \cite{yin2020,Ji2021} for \NF\ and \AF. As can be judged from Figs.\ \ref{fig:DOS_Na} and \ref{fig:DOS_Ag}, this is possible for both \NF\ and \AF\ in the hybrid DFT framework, and for \NF\ also in the PBEsol+U framework. We note, however, that this is rather precarious since, as has been mentioned above, varying $U_{eff}$ or $\alpha$ has multiple consequences, for instance in regards to the amount of Fe(d) contributions to the top of the valence band. 

In modelling the PM state we have chosen to use a \Ueff\ value of 3 eV, since this gives the same ordering of the unoccupied e$_{g}$/t$_{2g}$ states as hybrid DFT for reasonable values of $\alpha$. The ordering of these states is indeed potentially important for e.g.\ the transport properties of these systems since they provide very different effective electron masses. Indeed, the t$_{2g}$ states are very flat, while the e$_{g}$ bands have some curvature, corresponding to lower effective electron masses. 

As the recent interest for \AF\ and \NF\ have emerged in the context of their membership of the lead-free halide double perovskite class of materials, it is instructive to point out similarities and differences of their electronic structure to some prototypical members of this class. The two prototypical members of this class are Cs$_2$AgInCl$_6$ and Cs$_2$AgBi(Br/Cl)$_6$. Indeed, these systems both have relatively disperse conduction bands, very different from the highly localized Fe(d)-based states of, in particular, \NF. A  possible implications of this is the self-localization of excess electrons, i.e. the formation of small-electron polarons on \FeIII\ lattice sites. Initial exploratory calculations (not shown) indeed indicate that electron polaron formation is likely, but a comprehensive investigation into this topic is beyond the scope of this paper and is the subject of intended future work. 

As has been mentioned throughout the paper, the presence of HS \FeIII\ have implications in regards to the computational techniques that have be used in order to accurately model these systems. Indeed, we have demonstrated that careful considerations regarding the magnetic state and strong correlation effects have to be made in order for accurate results to be obtained. Such considerations are not normally required in modelling systems within the class of halide perovskites.

\section{\label{sec:conclusion}Conclusion}
To conclude we have studied the electronic structure of \FeIII containing double halide perovskites \NF\ and \AF. We have investigated the impact on varying the effective on-site Hubbard interaction , $U_{eff}$, in  DFT+U and the impact of the fraction of exact exchange, $\alpha$, in the hybrid-DFT framework. We find that while basic structural properties are relatively insensitive to the choice of $U_{eff}$ or $\alpha$, several details of the electronic structure vary strongly. In particular, we find a set of highly localized Fe(d) states in the electronic structure of both \NF\ and \AF, which vary sensitively with \Ueff\ and $\alpha$. A value of $U_{eff}$ = 3 eV in DFT+U yields the same ordering of the Fe t$_{2g}$ and e$_g$ states as in hybrid-DFT, for reasonable values of $\alpha$.

We have further investigated the impact of the magnetic state of these compounds. We find both systems to be antiferromagnetic at low T, but with rather small energy difference across different magnetic orderings. This is in agreement with the rather low experimental Ne\'el temperatures of ~3 K and ~18 K for \NF\ and \AF, respectively \cite{Xue2022}. Furthermore, we have revealed the effect of disordered magnetism on the electronic band structures of these systems, finding that while \NF\ is only weakly affected, \AF\ is substantially influenced by the disordered magnetic state.

\section{Acknowledgments}
This work was financially supported by Knut and Alice Wallenberg Foundation (Dnr. KAW 2019.0082). SSI and IAA acknowledge support from the Swedish Government Strategic Research Areas in Materials Science on Functional Materials at Linköping University (Faculty Grant SFO-Mat-LiU No. 2009-00971). S.I.S. acknowledges the support from Swedish Research Council (VR) (Project No. 2019-05551) and the ERC (synergy grant FASTCORR project 854843). The computations were enabled by resources provided by the Swedish National Infrastructure for Computing (SNIC), partially funded by the Swedish Research Council through grant agreement no. 2018-05973.

%\bibliography{main} % Produces the bibliography via BibTeX.

\end{document}